\documentclass[preprint2]{aastex}
\usepackage{spr-astr-addons}
\usepackage{latexsym}
\usepackage{epsfig} 
\usepackage{natbib}
\usepackage{amssymb,amsbsy}
\usepackage{amsmath}
\RequirePackage{color}
\newcommand{\barr}{\begin{array}}
\newcommand{\earr}{\end{array}}
\newcommand{\berr}{\begin{eqnarray}}
\newcommand{\err}{\end{eqnarray}}
\newcommand{\berrno}{\begin{eqnarray*}}
\newcommand{\errno}{\end{eqnarray*}}
\newcommand{\be}{\begin{equation}}

\newcommand{\pol}[1]{\stackrel{\rm LCP}{\mathrm{RCP}}}

\shorttitle{SEDs in UVIT}
\shortauthors{S. Ravichadran., Preethi Krishnamoorthy, Margarita Safonova and Jayant Murthy}
\begin{document}
\title{Large Scale Extinction Maps with UVIT }
\author{S.~Ravichandran\altaffilmark{1}}
\and
\author{K.~Preethi\altaffilmark{2}}
\affil{Christ University, Hosur Road,
Bangalore 560029, India}
\and
\author{M.~Safonova\altaffilmark{3}}
\and
\author{Jayant Murthy\altaffilmark{4}}

\affil{Indian Institute of Astrophysics, Koramangala 2nd block,
Bangalore 560034, India}

\altaffiltext{1}{s.ravichandran@christuniversity.in}
\altaffiltext{2}{preethi524@gmail.com}
\altaffiltext{3}{rita@iiap.res.in}
\altaffiltext{4}{jmurthy@yahoo.com}

\date{Received date / 
Accepted date} 

\abstract
 
The Ultraviolet Imaging Telescope (UVIT) is scheduled to be launched as a part of the
ASTROSAT satellite. As part of the mission planning for the instrument we have studied the 
efficacy of UVIT observations for interstellar extinction measurements. We find that in the
best case scenario, the UVIT can measure the reddening to an accuracy of about 0.02 magnitudes,
which combined with the derived distances to the stars, will enable us to model the 
three-dimensional distribution of extinction in our Galaxy. The knowledge of the distribution
of the ISM will then be used to study distant objects, affected by it. This work points the way
to further refining the UVIT mission plan to best satisfy different science studies.
 

\maketitle

\section{Introduction}

It is very difficult to model the three-dimensional distribution of interstellar matter (ISM) in our Galaxy 
because of the general lack of distance information. One of the few ways to probe the distance of the 
interstellar gas and dust is through absorption line measurements (in the case of gas) or through extinction 
(in the case of the dust), but these have been limited to a relatively small number of directions. This 
has impacted studies of extragalactic objects, not to mention the extragalactic background, because 
of the unknown effects of interstellar dust in the Milky Way.

The most reliable means of determining the amount of dust in a given line of sight is through observations 
of the extinction along the line of sight to stars across the sky. This originally involved comparing similar stars 
and measuring the difference between their spectra, but stellar models have now become sufficiently accurate 
so that they can be used for comparison instead, greatly extending the utility of this procedure 
\citep{FM2007}. Further, if the distance to a star is known, then, in principle, the distance to the 
actual scattering dust may be inferred. This method traditionally required spectral observations of 
moderate resolution and has been limited to those specific sight lines where such observations existed.

A much wider coverage was achieved by \citet{BH1982} who combined observations of 21 cm 
emission with the gas-to-dust ratio of \citet{BSD1978} to provide reddening maps over the sky. This 
was supplanted by \citet{SFD1998} who used the infrared emission from the Infrared Astronomy 
Satellite (IRAS) and the Cosmic Orbiting Background Explorer (COBE) to estimate the extinction over 
the entire sky. Both methods are model-dependent and, moreover, can only estimate the integrated 
dust column density. This can be overcome by using modern photometric surveys which observe large 
areas of the sky in multiple bands. Amongst these are \citet{Marshall2006}, who have used the Two 
Micron All Sky Survey (2MASS) to model the extinction in the Galactic plane, and \citet{Finkelman2008},
who have used SALT data combined with Sloan Digital Sky Survey Data (SDSS) to study extinction at 
high galactic latitudes. These surveys yield both the extinction along any line of sight and the distance 
to the background star and so may be used to develop a three-dimensional map of the dust in our Galaxy.

We report here on our plans to use data from the Ultraviolet Imaging Telescope (UVIT) aboard the 
ASTROSAT mission to probe the extinction across the entire sky. This mission has been 
in development since 2000 \citep{agrawal} and is expected to be launched on a Polar Satellite 
Launch Vehicle (PSLV) rocket by the Indian Space Research Organization (ISRO) in 2013. The 
UVIT instrument is being developed at the Indian Institute of Astrophysics (IIA) and includes three 
telescopes, two in the ultraviolet (FUV and NUV) and one in the visible, each including a filter wheel 
with a number of different filters \citep{kumar2012}. We have run a series of simulations to determine 
which filters are best suited for our purpose and will use these results in planning our observations 
with the UVIT.

\section{Data and Simulations}

\begin{figure*}[ht!]
\begin{center}
\includegraphics[scale=0.5,angle=-90]{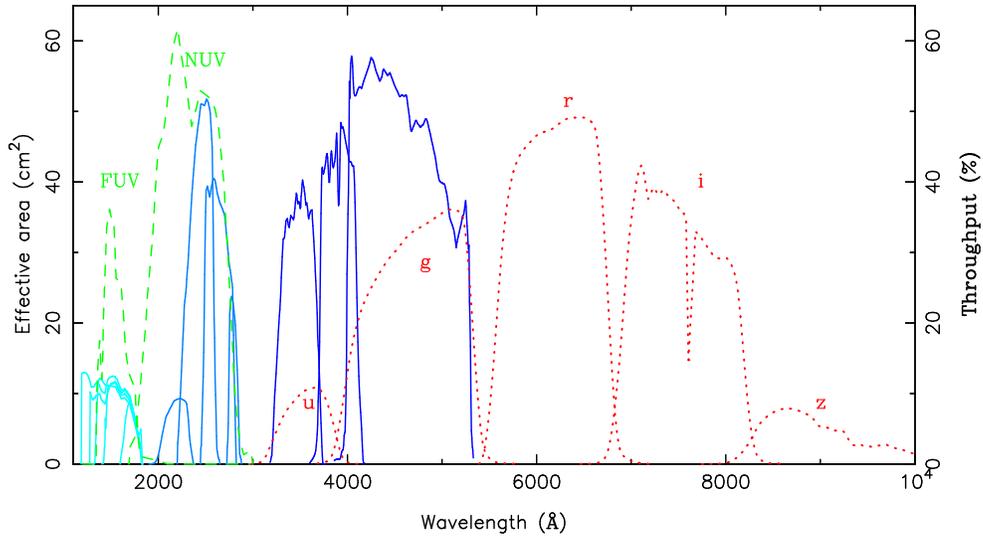}
\caption{In this plot we have combined the effective area curves of the UVIT filter set: FUV filters in cyan,
NUV filters in light blue, visible filters in dark blue. We also show for comparison GALEX effective area curves 
(green dashed) and the throughput functions of SDSS filter set (red dotted). Note that SDSS curves are
given as a throughput function in \%, not as effective areas.}
\label{fig:filters}
\end{center}
\end{figure*}

The standard operating procedure of UVIT will be to observe selected targets with a nominal exposure time 
of about 1000 seconds. Although, due to the multi-wavelength nature of the ASTROSAT mission (there are 
four other instruments on board, covering the entire X-ray range, apart from the UVIT) and the need to 
accommodate all the instruments in the time planning, there will be a number of different operating modes.
The standard data products will consist of FITS binary tables with a time ordered  photon list, FITS images 
of each field, and a catalog of astrometrically and photometrically corrected point sources. We will start our 
work with that catalog, and match the UVIT sources with SDSS and GALEX sources. The UVIT filter set 
covers a broad range of UV to visible spectral region from about 1200 to 5500 \AA\, dividing the range 
into 12 narrow and broad bands. We considered 5 filters in the FUV telescope (CaF$_2$\_1, CaF$_2$\_2 
and BaF$_2$, Silica and Sapphire), 4 filters in the NUV channel (NUVN2, NUVB4, NUVB13, NUVB15) 
and 3 filters in the visible channel (VIS1, VIS2 and VIS3). The effective areas of these filters, together 
with GALEX and SDSS bands, are shown in Fig.~\ref{fig:filters}. These filters were chosen for their 
scientific utility for different programs, with one (NUVB15) selected especially to study the 
2175 \AA \,extinction bump. However, because the visible imager was chosen primarily for registration 
purposes and the data may not be complete or photometrically accurate, we have used only the UV 
filters in this work. Combining these with the two GALEX bands and the five SDSS bands, we will have a 
total of 16 photometric magnitudes from the FUV through to the near infrared (1400 -- 9000 \AA), 
if the data are available in all bands (Table 1).

\begin{table}[h!]
\caption{Effective wavelengths and predicted $m_{\rm AB}$ in different filters for a $B0V$ type star.}
\begin{center}
\begin{tabular}{lcl}
\hline
Filter  &   	Effective Wavelength (\AA)   &  $m_{\rm AB}$    \\
\multicolumn{3}{c}{$B0V$ star, $T = 30000; E(B-V) = 0$}     \\
\hline 
GALEX FUV		&   	1511.2     & 3.145  \\
GALEX NUV		&   	2203.1    & 3.418 	 \\
UVIT CaF$_2$\_1	&   	1483.5    & 3.122 \\
UVIT CaF$_2$\_2	&   	1487.5    & 3.126 \\
UVIT BaF$_2$ 	&   	1514.7    & 3.147 	 \\
UVIT Sapphire	  	&   	1591.4    & 3.189 	\\
UVIT Silica	 	&   	1701.9    & 3.181 	\\
UVIT NUVB15	  	&   	2175.7    & 3.347 	 \\
UVIT NUVB13	 	 &   	2421.0    & 3.47 \\
UVIT NUVB4	  	&   	2618.1    & 3.576 	 \\
UVIT NUVN2	 	 &   	2790.9    & 3.656 	\\
SDSS $u$			&   	3515.7    & 4.02 \\
SDSS $g$ 		&    4568.5    & 4.352 	 \\
SDSS $r$ 		&   	6093.5   & 4.854 	\\
SDSS $i$			&   	7048.5    & 5.224 \\
SDSS $z$			&   	8833.7    & 5.57 	 \\
\hline
\end{tabular}
\end{center}
\label{tab:table1}
\end{table}

\begin{figure}[h!]
\begin{center}
\includegraphics[scale=0.42]{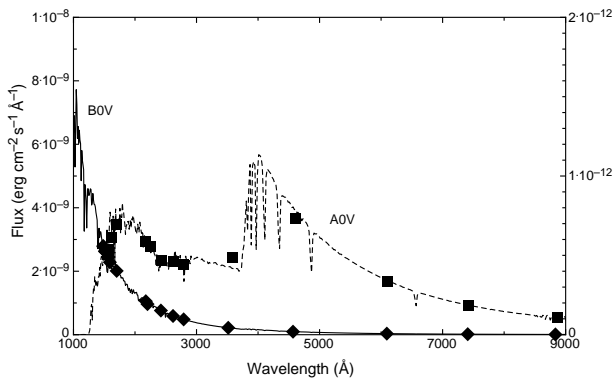}
\caption{The stellar spectra of an unreddened B0V star (solid line) and an unreddened A0V star 
(dashed line -- right axis) are plotted with diamonds and squares marking respectively the 
effective flux in each of the GALEX FUV (at 1557 \AA) and NUV bands (at 2261 \AA) in red, 
9 UVIT bands in green and 5 SDSS filters in blue. Note that the effective wavelength is dependent 
 on both the stellar spectrum and the filter response.}
\label{fig:fig1}
\end{center}
\end{figure}

\begin{figure}[h!]
\begin{center}
\includegraphics[scale=0.45]{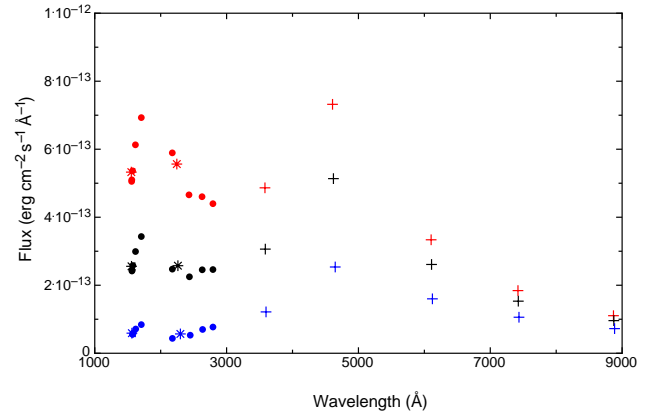}
\caption{The effective flux in each of the filters for an A0V star with $E(B-V) = 0$ (red); 
$E(B-V) = 0.1$ (black); 
and $E(B-V) = 0.3$ (blue). Asterisks represent the flux in the two GALEX bands, circles -- in the UVIT 
bands and crosses -- in the SDSS bands.}
\label{fig:fig2}
\end{center}
\end{figure}

We will compare the model spectral energy distributions (SEDs) with data observed by the UVIT. 
In this work, we have created the model SEDs for stars of different spectral types reddened by 
interstellar dust. The stellar SEDs are based on the model spectra of \citet{Castelli2004}, provided 
as fluxes as a function of temperature. Each spectrum was reddened by an extinction curve, 
based on the standard Milky Way curve of \citet{Draine2003}, and then convolved with the all filter 
curves to produce a magnitude for each of the 16 bands. This is illustrated in Fig.~\ref{fig:fig1}, 
where we have plotted the spectra of an unreddened B0V and an unreddened A0V star with 
effective fluxes in each of the bands. Note that the points do not always fall on the stellar spectrum 
because the integration was done over the entire filter band. Fig.~\ref{fig:fig2} shows the effective 
fluxes for an A0V star for three different values of the interstellar reddening. 

\begin{figure}[h!]
\begin{center}
\includegraphics[scale=0.45]{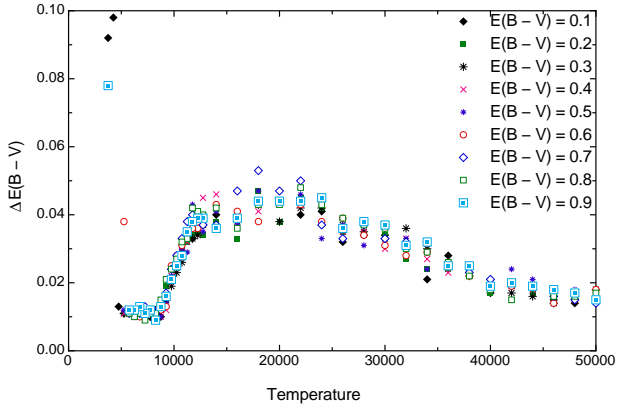}
\caption{Standard deviations in the $E(B - V)$ derived from Monte Carlo simulations 
of the observed fluxes through the GALEX, UVIT and SDSS bands.}
\label{fig:fig3}
\end{center}
\end{figure}

We have performed Monte Carlo simulations to estimate the uncertainty expected from the UVIT 
observations. Each simulation consisted of a series of 100 independent runs for a 20th magnitude 
star, with different combinations of temperature and $E(B - V)$ with an assumed error of 0.1 
magnitudes in each of the bands. The deviations of the 
derived $E(B - V)$ from the actual values are plotted in Fig.~\ref{fig:fig3} as a function of the stellar 
temperature. We are best able to reproduce the original values for A and F stars, with a deviation in 
$E(B - V)$ of 0.02 magnitudes with an upper limit of 0.05 magnitudes, for stellar temperatures of 
about 20,000 K. The cooler stars (G type and later) have no detectable UV emission and, hence, 
are not detectable with the UVIT.

\begin{figure}[h!]
\begin{center}
\includegraphics[scale=0.45]{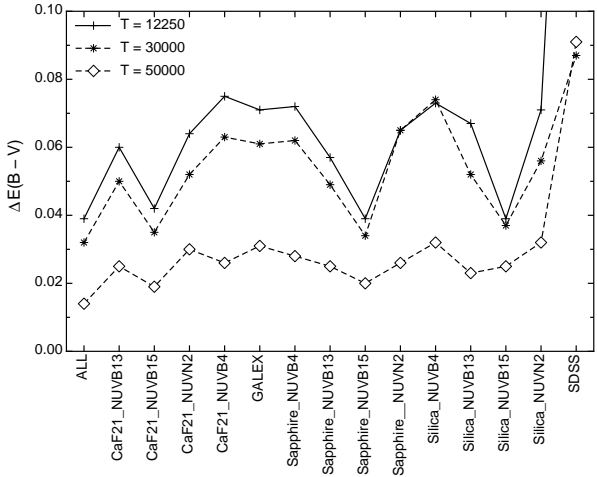}
\caption{The deviations in the derived E(B-V) are shown for each of the different filter 
combinations. ALL means that all filters were used (see Fig.~\ref{fig:fig1}). Three FUV
filters, CaF$_2$\_1, CaF$_2$\_2 and BaF$_2$ are quite similar (see Fig.~\ref{fig:filters}) 
and we show here the results for only one, CaF$_2$\_1. Note that the SDSS data alone 
cannot constrain the reddening effectively.}
\label{fig:fig4}
\end{center}
\end{figure}

The best possible case would be if we had multiple observations 
of the same field through each of the combinations of FUV and NUV filters. However, 
in most cases, we will not have multiple observations of any given location, and thus we have to optimize 
the choice of filters to maximize the science output, which implies, in this case, the minimization of 
the $\Delta E(B - V)$. We 
have simulated different combinations of FUV and NUV filters, and calculated the standard deviations 
for different spectral types and reddening values. The standard deviations for three different spectral types 
are shown in Fig.~\ref{fig:fig4}, assuming a reddening ($E(B - V)$) of 0.5 magnitudes. In all cases, 
the best fit to the data (the lowest deviations) are for the NUVB15 filter in the UVIT NUV telescope. 
We are much less sensitive to the FUV telescope, where any of the choices are acceptable.

\section{Conclusions}

We have begun the process of observational mission planning for the UVIT instrument by 
calculating the flux in each of the UVIT bands to maximize the science. In this work, 
we describe our plans to make a three-dimensional dust map over the sky using the UVIT 
point source catalog, combined with GALEX and SDSS observations. We have determined the 
best filter combination for this extinction work as NUVB5 filter with any of the the FUV filter, 
but since observations are generally serendipitous in nature, we will use all the UVIT 
data for our purpose. The process of observational mission planning for UVIT will 
include the optimization of filters combinations for all scientific objectives. For 
example, we will calculate SEDs for different scientific programs (such as, for ex., 
stellar astrophysics, observations of globular clusters, AGN, etc.) and select the optimal filters.

\nocite{*}
\bibliographystyle{spr-mp-nameyear-cnd}
\bibliography{biblio-u1}

\end{document}